\documentclass[10pt]{article}
\usepackage{graphicx}

\usepackage{lineno}

\usepackage{comment}

\usepackage{hyperref}
\hypersetup{
    colorlinks=true,
    linkcolor=blue,
    filecolor=blue,      
    urlcolor=blue,
}

\def\Title#1{\begin{center} {\Large #1 } \end{center}}
\def\Author#1{\begin{center}{ \sc #1} \end{center}}
\def\Address#1{\begin{center}{ \it #1} \end{center}}

\newcommand\pubblock{\rightline{\begin{tabular}{l} Proceedings of DPF 2017\\ \pubnumber\\
         \pubdate  \end{tabular}}}

\newenvironment{Abstract}{\begin{quotation} \begin{center} 
             \large ABSTRACT \end{center}\bigskip 
      \begin{center}\begin{large}}{\end{large}\end{center} \end{quotation}}

\newenvironment{Presented}{\begin{quotation} \begin{center} 
         \end{center}\bigskip 
      \begin{center}\begin{large}}{\end{large}\end{center} \end{quotation}}

\def\Acknowledgements{\bigskip  \bigskip \begin{center} \begin{large}
             \bf ACKNOWLEDGEMENTS \end{large}\end{center}}

\def\beq{\begin{equation}}
\def\eeq#1{\label{#1}\end{equation}}
\def\eeqn{\end{equation}}

\def\beqa{\begin{eqnarray}}
\def\eeqa#1{\label{#1}\end{eqnarray}}
\def\eeqan{\end{eqnarray}}

\let\bar=\overbar

\def\Dslash{\not{\hbox{\kern-4pt $D$}}}
\def\dslash{\not{\hbox{\kern-2pt $\del$}}}

\def\msb{{\bar{\ssstyle M \kern -1pt S}}}

\usepackage{color}
\usepackage{cite}

\newcommand\pubnumber{ ATL-INDET-PROC-2017-003 }

\newcommand\pubdate{\today}

\def\affiliation{
On behalf of the ATLAS Collaboration, \\
Physics Division, Lawrence Berkeley National Laboratory\\
Berkeley, CA 94720, USA}

\begin{document}

\large
\begin{titlepage}
\pubblock

\vfill
\Title{  Modeling Radiation Damage to Pixel Sensors in the ATLAS Detector }
\vfill

\Author{ Benjamin Nachman  }
\Address{\affiliation}
\vfill
\begin{Abstract}

Silicon pixel detectors are at the core of the current ATLAS detector and its planned upgrade. As the detectors in closest proximity to the interaction point, they will be exposed to a significant amount of radiation: prior to the HL-LHC, the innermost layers will receive a fluence in excess of $10^{15}$ 1 MeV $n_\mathrm{eq}/\mathrm{cm}^2$ and the HL-LHC detector upgrades must cope with an order of magnitude higher fluence integrated over their lifetimes. This talk presents a digitization model that includes radiation damage effects to the ATLAS Pixel sensors for the first time. After a thorough description of the setup, predictions for basic pixel cluster properties are presented alongside first validation studies with Run 2 collision data.

\end{Abstract}
\vfill

\begin{Presented}
Talk presented at the APS Division of Particles and Fields Meeting (DPF 2017) July 31-August 4, 2017, Fermilab. C170731\\ 
\end{Presented}
\vfill
\end{titlepage}
\def\thefootnote{\fnsymbol{footnote}}
\setcounter{footnote}{0}
%

\normalsize 


\section{Introduction}

\begin{figure}[htb]
\begin{center}
\includegraphics[width=0.8\textwidth]{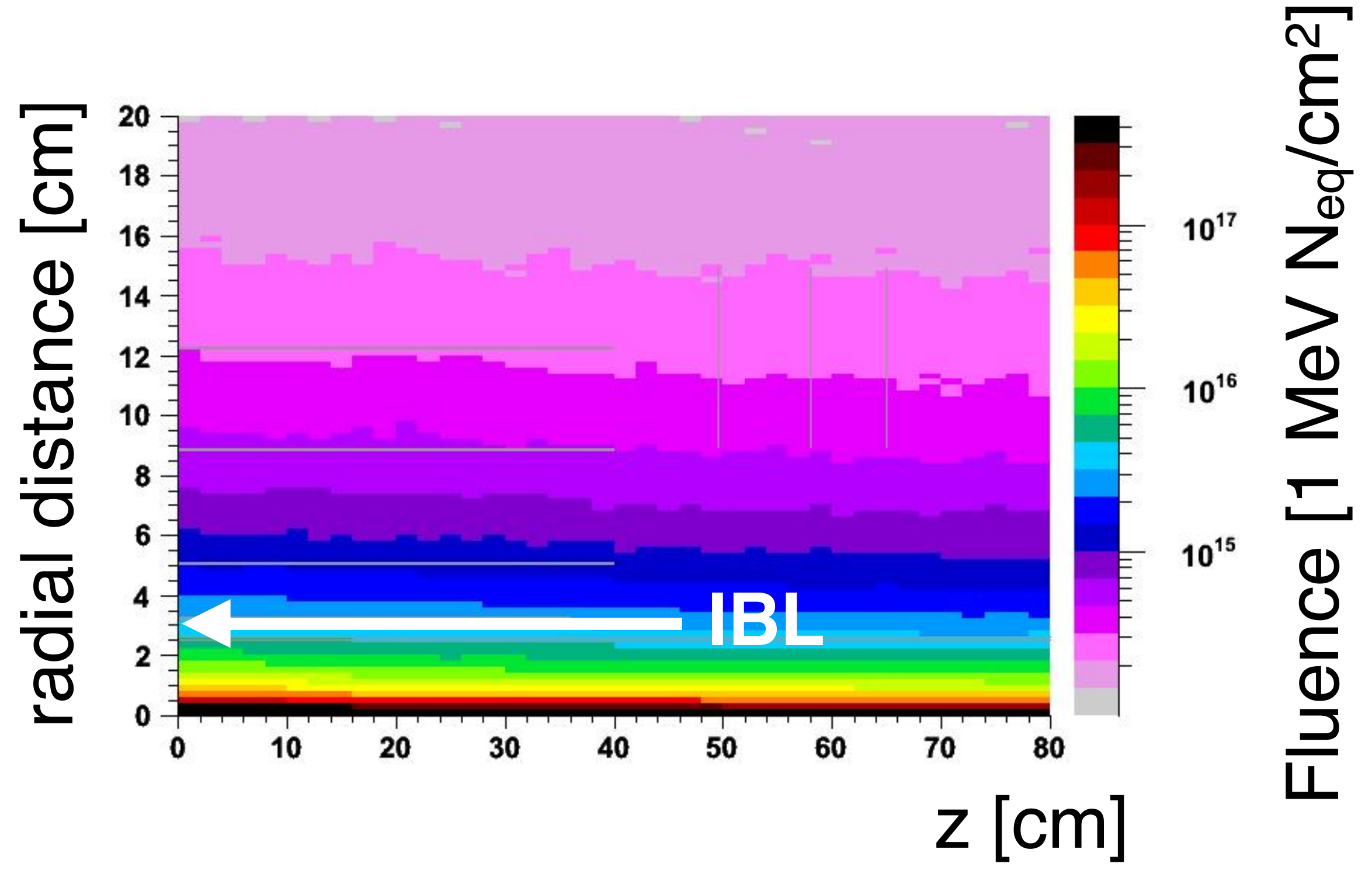}
\caption{The predicted fluence as a function of longitudinal and radial distance from the geometric center of the ATLAS detector after approximately the end of the LHC using the particle transport code FLUKA~\cite{Battistoni:2007zzb,Ferrari:898301}.   Plot adapted from the technical design report, Ref.~\cite{Capeans:1291633}.}
\label{fig:fig1}
\end{center}
\end{figure}

One critical side-effect of a high proton-proton collision rate at the Large Hadron Collider (LHC) is extreme radiation damage to the detectors closest to the interaction points.  In the ATLAS detector~\cite{Detector}, the closest subsystem is the pixel detector~\cite{Aad:2008zz,Capeans:1291633}.  The pixel detector is composed of four barrel layers and multiple disks, allowing for charged particle reconstruction out to $|\eta|=2.5$ and covering the full azimuthal angle $\phi\in[0,2\pi]$.  The innermost barrel pixel layer, the Insertable B-Layer (IBL)~\cite{Capeans:1291633}, is 3.3 cm from the interaction point and is composed of 200 $\mu$m thick planar and 230 $\mu$m thick 3D sensors with a $50\times 250$ $\mu$m$^2$ pitch.  The IBL was inserted during the long LHC shutdown between Runs 1 and 2\footnote{The LHC run plan includes 5 fb$^{-1}$ of $\sqrt{s}=7$ TeV and 20 fb$^{-1}$ of $\sqrt{s}=8$ TeV in Run 1, approximately 100 fb$^{-1}$ in Run 2 at $\sqrt{s}=13$ TeV, and an additional 200 fb$^{-1}$ by the end of Run 3.}; it has already accumulated over $2\times 10^{14}$~1~MeV~$\mathrm{n}_\mathrm{eq}/\mathrm{cm}^2$ of non-ionizing energy loss (NIEL).  This NIEL is about the same dose that the innermost layer of the original pixel detector has absorbed and about twice (four times) the amount on the third (fourth) layer.  This \textit{fluence} introduces defects in the sensors' silicon lattice which in turn has detrimental effects on detector performance.  In particular, the electric field is deformed (also the depletion voltage increases) and ionized charge can be trapped before being collected.  This talk reported on current measurements of radiation damage effects in the present pixel detector as well as the introduction of a new simulation that incorporates radiation damage for the first time for the ATLAS pixel detector.  Monitoring and modeling radiation damage will continue to become even more important as the current innermost layer(s) exceed $1\times 10^{15}$~1~MeV~$\mathrm{n}_\mathrm{eq}/\mathrm{cm}^2$ by the end of the LHC (Runs 1-3) and as the upgraded inner tracker upgrade (ITk) sensors accumulate in excess of $10^{16}$~1~MeV~$\mathrm{n}_\mathrm{eq}/\mathrm{cm}^2$ for the high-luminosity LHC (HL-LHC) (Runs 4 and 5 with 3 ab$^{-1}$).  Figure~\ref{fig:fig1} indicates the predicted fluence as a function of longitudinal and radial distance from the geometric center of the ATLAS detector after approximately the end of the LHC in 2023.  All pixel layers will reach or exceed their design fluence limit.

\clearpage

\section{Measurements of Radiation Damage}

\begin{figure}[htb]
\begin{center}
\includegraphics[width=0.85\textwidth]{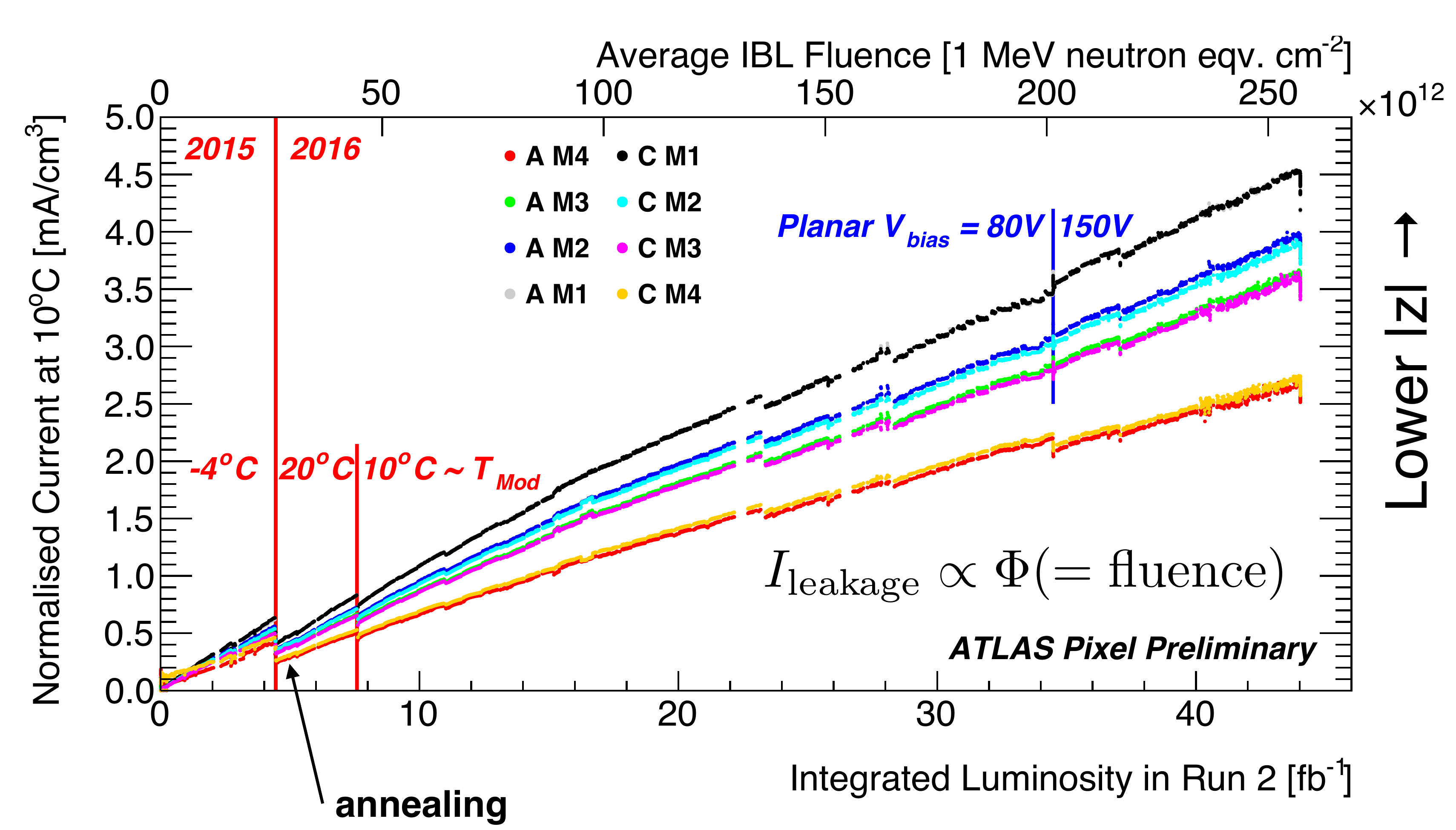}
\caption{The leakage current measured on the IBL as a function of integrated luminosity collected during the 2015 and 2016 runs.  The IBL staves are divided into $A$ and $C$ sides depending on their direction from the interaction point.  There are four groupings of modules on each side, with M1 closest to the detector center (small $|z|$) and M4 furthest away (large $|z|$).  Module groups M1-M3 use planar sensors while M4 is composed of modules with 3D sensors.  Plot adapted from Ref.~\cite{PIX-2016-006}.}
\label{fig:fig2}
\end{center}
\end{figure}

Already with the Run 1+2 dataset, there are clear indications of radiation damage in the ATLAS silicon pixels.  One of the most important diagnostics for evaluating the radiation damage is the sensor leakage current.  Silicon crystal defects can be viewed as energy levels in the band gap; more defects result in more thermal charge carriers and thus an increased current even in the absence of an ionizing particle flux.  The number of defects is proportional the fluence $\Phi$ and therefore the leakage current $I_\mathrm{leakage}\propto \Phi$. Figure~\ref{fig:fig2} shows the leakage current measured on the IBL as a function of integrated luminosity collected during the 2015 and 2016 runs.  Since the fluence is proportional to the integrated luminosity, the leakage current increases nearly linearly with the integrated luminosity.  One complication is \textit{annealing} in which defects can decay with time.  This is relevant when the detector is warm as is typical during the shutdown between runs.  If all IBL modules were exposed to the same fluence, the slope of the leakage current curves in Figure~\ref{fig:fig2} would be the same.  The differences between module groups is an indication of a $|z|$-dependent irradiation profile, with lower fluences further away from the interaction point.  

\begin{figure}[htb]
\begin{center}
\includegraphics[width=0.7\textwidth]{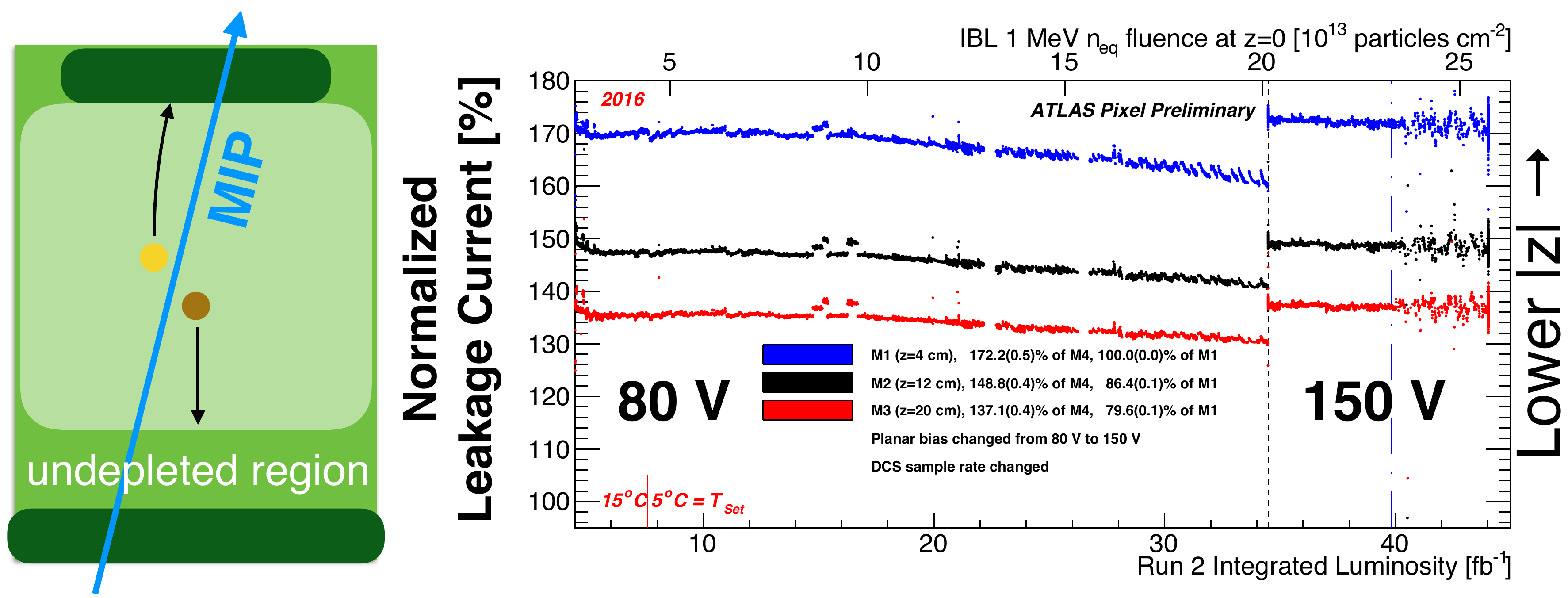}
\caption{The ratio of the leakage current in the M1-M3 planar pixel IBL module groups normalized by the M4 3D pixel IBL module current as a function of integrated luminosity.  Plot adapted from Ref.~\cite{PIX-2016-006}. The left plot is a schematic diagram to illustrate an under-depleted sensor.}
\label{fig:fig3}
\end{center}
\end{figure}

The leakage current is important for noise and power consumption but does not directly influence charge collection.  The remainder of this section describes the impact of radiation damage on three observables that each directly contribute to the degradation of tracking performance.  The first is the depletion volume.  Electron-hole pairs generated by ionizing radiation can only be collected if they are produced in the depletion region so that they drift in the presence of the electric field.  There are several methods for measuring the depletion volume; one way is to measure the leakage current relative to a reference.  As described earlier, the leakage current is proportional to fluence; it is also proportional to the depletion volume.   Figure~\ref{fig:fig3} shows the leakage current in the planar pixel modules of the IBL relative to the M4 modules that have 3D sensors.  The depletion voltage for the 3D sensors is very low and is always depleted over the dataset shown in Figure~\ref{fig:fig3}.  The vertical separation between lines is the same $|z|$-dependence of the fluence that was shown in Figure~\ref{fig:fig2}.  The ratio is nearly flat until about $\int\mathcal{L} dt\approx 15$ fb$^{-1}$, where it starts to decrease.  This decrease results since the depletion volume shrinks to less than the whole sensor.  After about $\int\mathcal{L} dt= 35$ fb$^{-1}$, the high voltage on the IBL was increased from 80 V to 150 V and the level and flatness are restored.

\begin{figure}[htb]
\begin{center}
\includegraphics[width=0.7\textwidth]{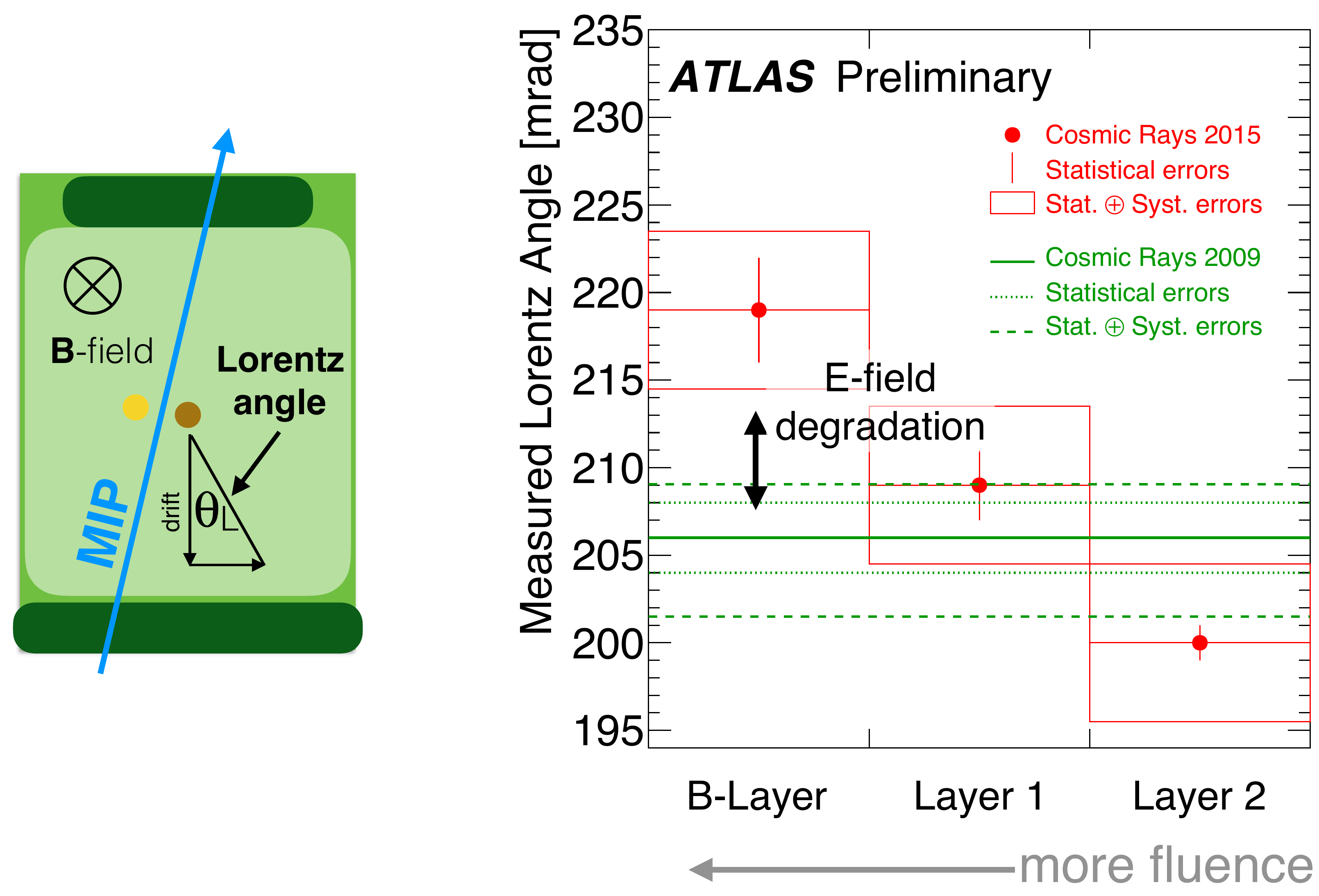}
\caption{ The measured Lorentz angle just before the Run 1 (2009) and Run 2 (2015) data taking using cosmic muons for the original three pixel layers. Plot adapted from Ref.~\cite{IDTR-2017-002}. The left plot is a schematic diagram to illustrate the Lorentz angle.}
\label{fig:fig5}
\end{center}
\end{figure}

Another important diagnostic for fluence-induced effects is the Lorentz angle.  In planar sensors, electrons and holes resulting from ionizing radiation drift at an angle with respect to the electric field direction as a result of a 2 T magnetic field that is perpendicular to the depth.  This Lorentz angle is proportional to the carrier mobility, which increases as the electric field degrades in the center of the sensor due to radiation damage.  Figure~\ref{fig:fig5} shows the measured Lorentz angle for the original three layers of the ATLAS pixel detector just before Run 1 and just before Run 2.  At the beginning of the LHC, all layers have the same Lorentz angle of about 205 mrad.  However, after all of Run 1, there is a significant increase in the Lorentz angle on the innermost of the three original pixel layers ($b$-layer) which has absorbed about $10^{14}$ 1 MeV $\mathrm{n}_\mathrm{eq}/\mathrm{cm}^\mathrm{2}$.

\begin{figure}[htb]
\begin{center}
\includegraphics[width=0.7\textwidth]{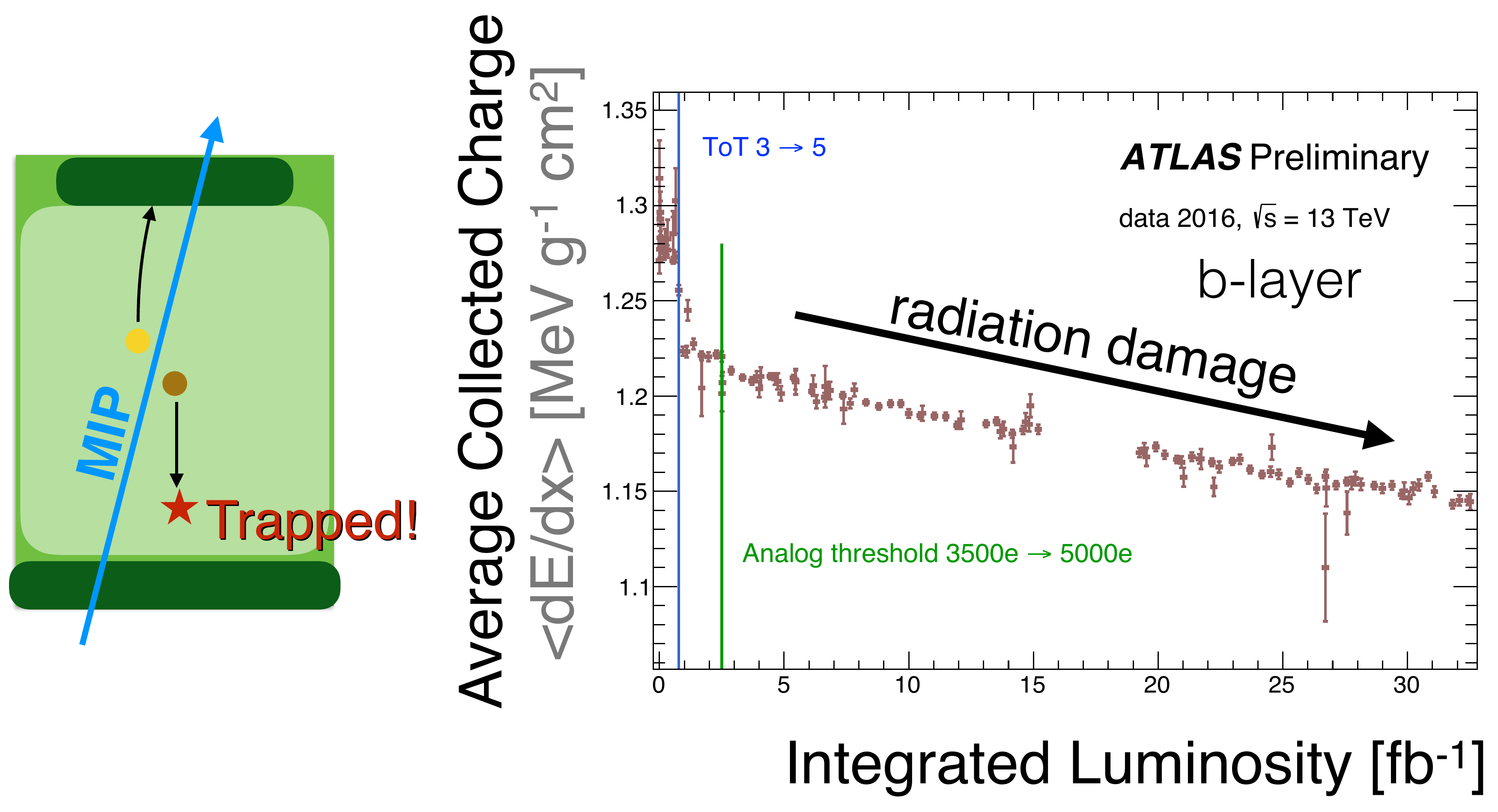}
\caption{The average collected charge on the $b$-layer for reconstructed high momentum tracks.  The jump after about 1 fb$^{-1}$ is from the increase in the digital threshold from 3 to 5; a smaller drop after about 2.5 fb$^{-1}$ is from an increase in the analog threshold from 3.5 ke to 5 ke.  The charge is built from the measured ToT per pixel and is corrected for the incidence angle that changes the path length. Plot adapted from Ref.~\cite{IDTR-2017-003}. The left plot is a schematic diagram to illustrate charge trapping.}
\label{fig:fig4}
\end{center}
\end{figure}

One final and direct probe of radiation damage is the measured charge deposited by a MIP traversing a single pixel layer.  The amount of trapping is proportional to the fluence and so it is expected that the collected charge decrease linearly with integrated luminosity.  This is demonstrated for the $b$-layer in Figure~\ref{fig:fig4}.  Aside from jump in the charge early in the run due to changes in thresholds, the average collected charge linearly decreases with integrated luminosity.  

The ATLAS pixel simulation currently does not include radiation damage effects, yet this section has already shown measurable degradation that will only continue with more fluence.  Simulating radiation damage is therefore critical in order to make accurate predictions for current and future detector performance that will enable searches for new particles and forces as well as precision measurements of Standard Model particles such as the Higgs boson.   The next section presents a digitization model that includes radiation damage effects to the ATLAS pixel sensors for the first time and considers both planar and 3D sensor designs.



\clearpage

\section{Simulation of Radiation Damage}

\begin{figure}[htb]
\begin{center}
\includegraphics[width=0.7\textwidth]{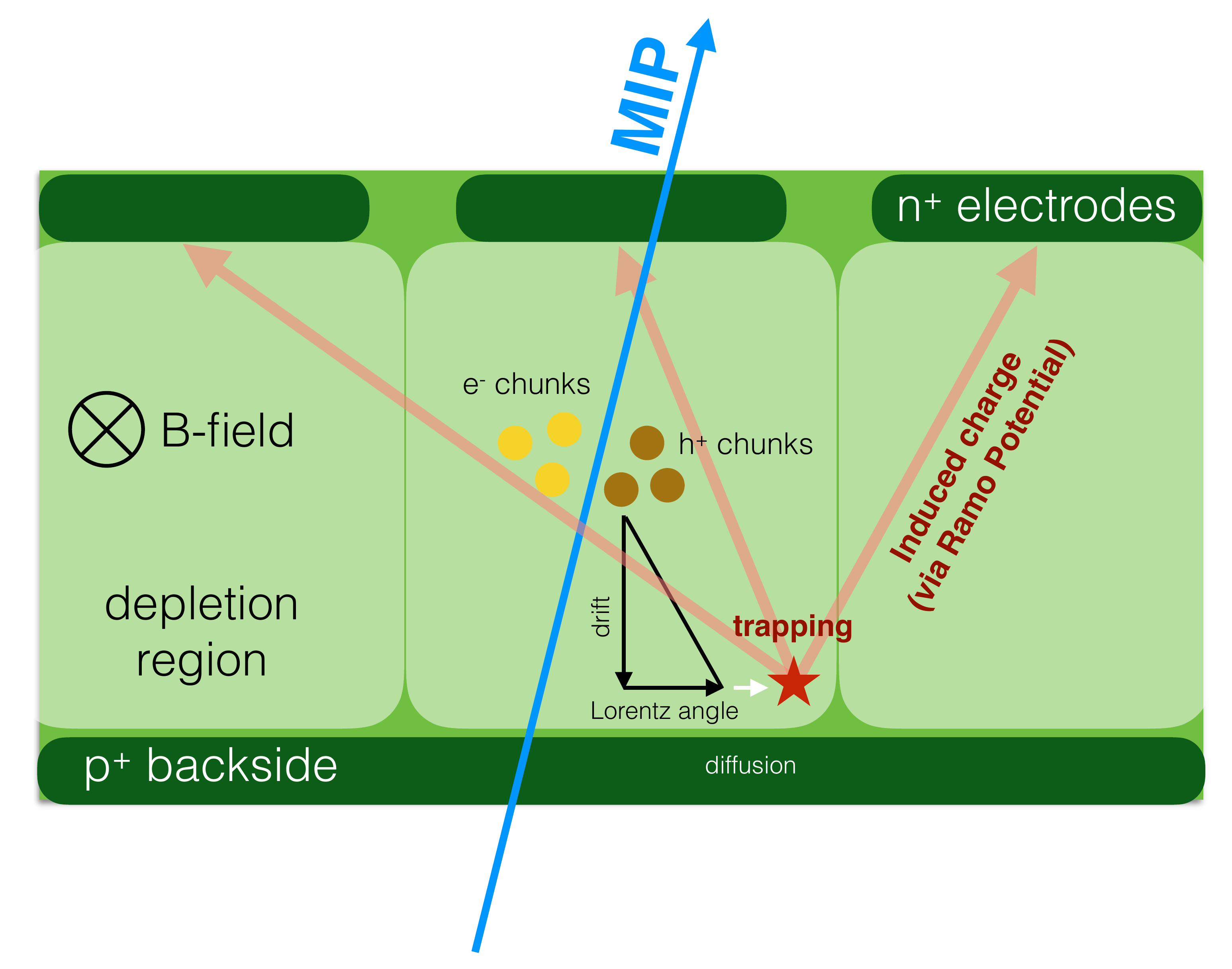}
\caption{A schematic overview of the physical processes included in the new digitization model.}
\label{fig:fig6}
\end{center}
\end{figure}

A model of radiation damage effects is included in the \textit{digitization} step of the ATLAS pixel simulation.  In this step, energy depositions in the silicon bulk are converted to digital signals using the Time Over Threshold (ToT) that are then used for track reconstruction as if they were real data.  Radiation damage effects are included in the simulation in two ways: the electric field is deformed and ionized charges can be trapped before being collected.    Figure~\ref{fig:fig6} presents an overview of the physical effects included during the charge collection portion of digitization.  The deformation in the electric field distorts charge transport including drift and diffusion.  Trapping reduces the overall collected charge, but the induced charge prior to trapping must also be accounted for, also on neighboring pixels.  Each of these components is described in more detail below.

\begin{figure}[htb]
\begin{center}
\includegraphics[width=0.7\textwidth]{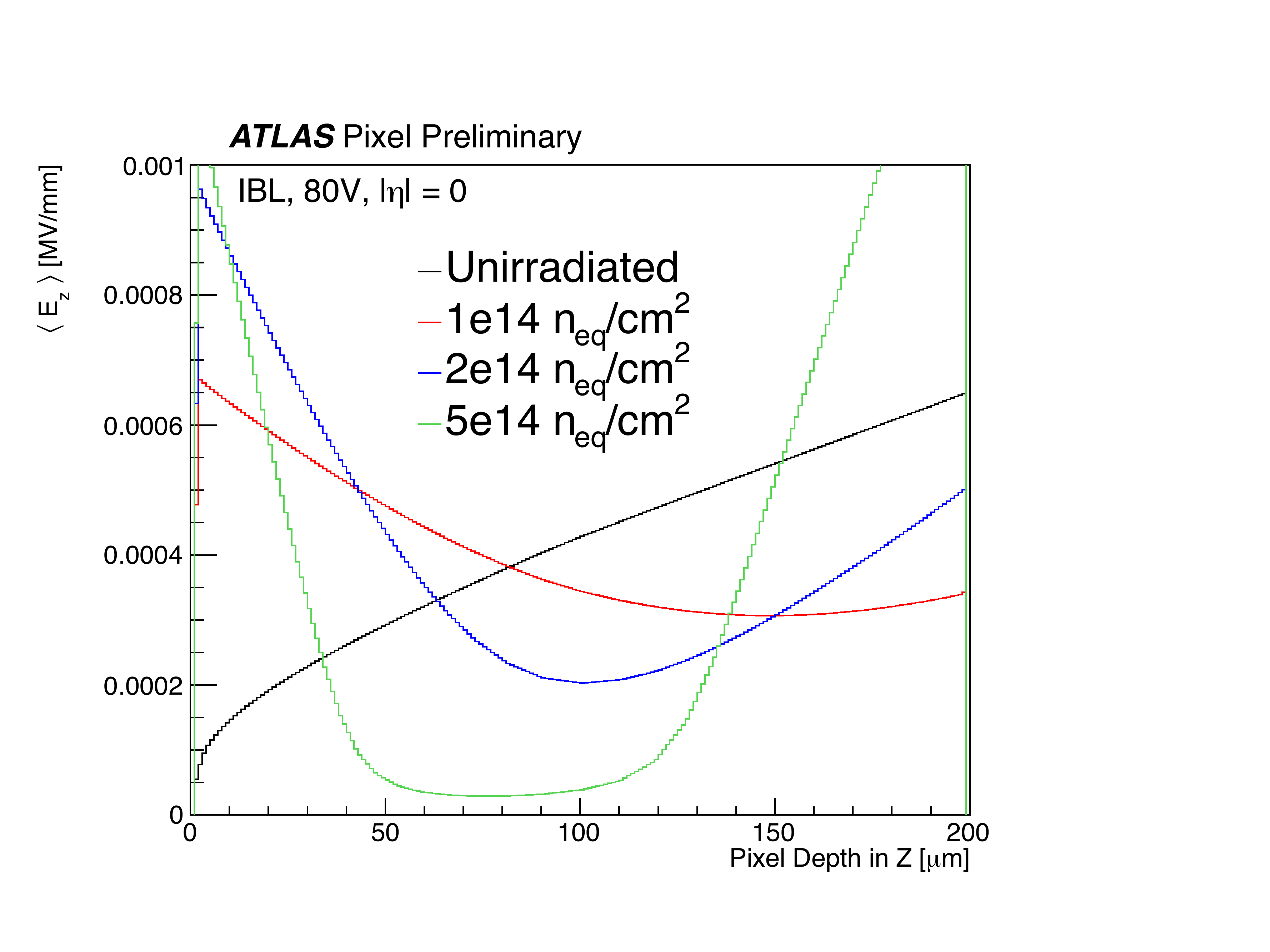}
\caption{The electric field profile predicted by the TCAD implementation of the Chiochia model for an IBL sensor at 80 V.  The pixel implant is on the left and the backside is on the right.  }
\label{fig:fig8}
\end{center}
\end{figure}

Fluence-dependent distortions in the electric field are calculated numerically using Technology Computer Aided Design (TCAD).  There are a variety of levels in the band gap from radiation-induced defects~\cite{Moll:1999kv}.  However, in practice it is too computationally expensive to include all levels and thus either two (for planar) or three (for 3D) effective traps are included in the TCAD models.  Planar sensors ($n$-type) are modeled using the Chiochia model~\cite{Chiochia2} which has one acceptor and one donor level.  3D sensors ($p$-type) are modeled using the Perugia model~\cite{7542192} with two acceptor and one donor levels.  The levels are specified by their activation energy, the electron and hole capture cross-sections, and the introduction rates.  For planar sensors, the field is nearly independent of the transverse $x$ and $y$ positions so only the $z$-dependence is used by the digitizer.  In contrast, for 3D sensors, the field is nearly independent of $z$ so only the $x$- and $y$-dependence is retained.  Only planar sensors are considered.  Figure~\ref{fig:fig8} shows the planar electric field as a function of $z$ for four fluences and a fixed high voltage of 80 V.  Prior to irradiation, the field is nearly linear and grows toward the backside.  After space-charge sign inversion, the field grows toward the pixel implant.  For even higher fluences, a double peak structure develops with a region of low field in the middle of the sensor.

\begin{figure}[htb]
\begin{center}
\includegraphics[width=0.7\textwidth]{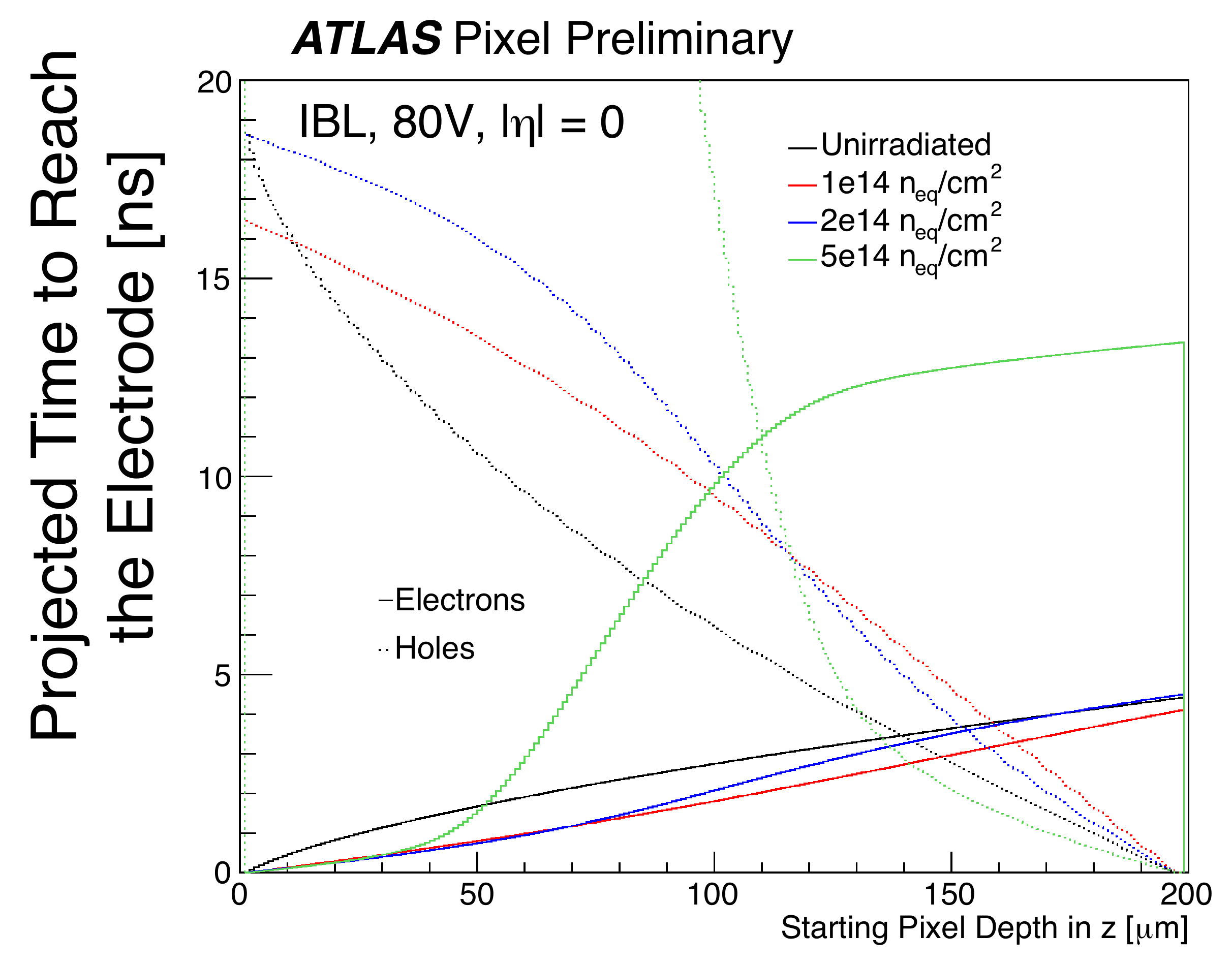}
\caption{The projected time to reach the charge carrier final destination as a function of starting depth inside a planar IBL sensor. The pixel implant is on the left and the backside is on the right. Trapping times are similar for electrons and holes and are exponentially distributed with mean values of about 30, 15, and 7 ns for $1$, $2$, and $5\times 10^{14}$ $n_\mathrm{eq}/\mathrm{cm}^2$, respectively~\cite{trapping1,trapping2}.}
\label{fig:fig9}
\end{center}
\end{figure}

In order to save time during digitization, many quantities which depend only on the electric field are pre-computed and stored in memory.  One such quantity is the time required to reach the electrode.  To determine if a particular electron or hole is trapped, the time to reach the electrode is compared to a random trapping time that depends on the fluence.  If it exceeds the trapping time, the charge carrier is declared trapped.  The time depends on the electric field and mobility $\mu$ via $t_\mathrm{electrode}=\int_{z_0}^{z_\mathrm{final}}dz/(\mu(E(z))E(z))$, where $z_0$ is the starting depth and $z_\mathrm{final}$ is $0$ for electrons and $200$ $\mu$m for holes.  Figure~\ref{fig:fig9} shows the fluence-dependence of the time to reach the electrode for the same fixed high voltage as Figure~\ref{fig:fig8}. The mobility of holes is significantly lower than for electrons so it takes much longer to reach their final destination.  As will be discussed below, most of the measured charge is generated by electrons, which minimizes the sensitivity to trapping.  3D sensors are designed to reduce the travel time to the electrode, further reducing the impact of charge trapping.

\begin{figure}[htb]
\begin{center}
\includegraphics[width=0.7\textwidth]{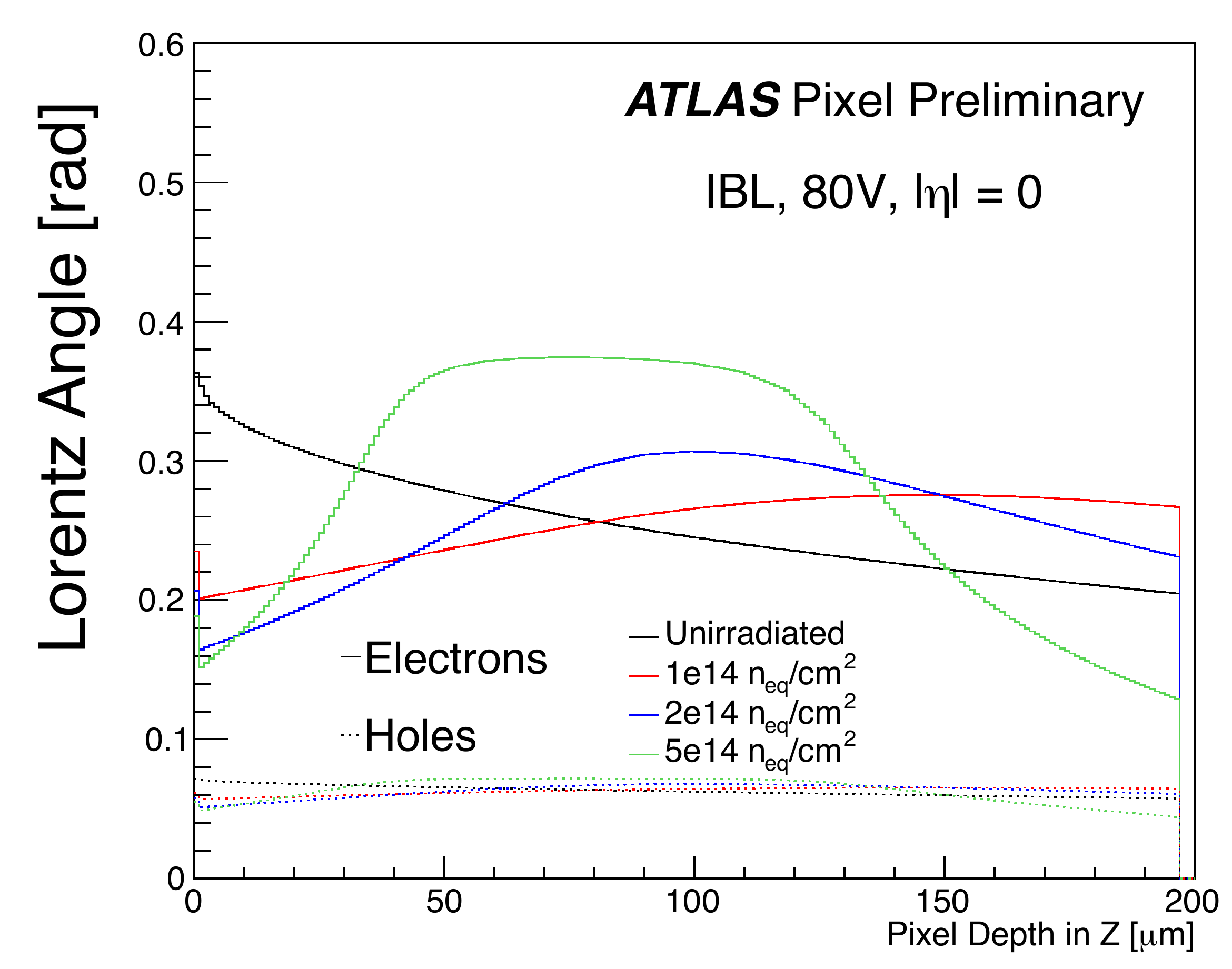}
\caption{The total Lorentz angle after integrating along the charge carrier path, as a function of the sensor depth for a variety of fluences. The pixel implant is on the left and the backside is on the right.}
\label{fig:fig10}
\end{center}
\end{figure}

Another quantity that can be pre-computed is the Lorentz angle.  As stated earlier, this angle is nearly proportional to mobility which itself depends on the field.  Therefore, the angle develops a fluence-dependence as demonstrated in Figure~\ref{fig:fig10}.  The electric field is lowest and thus the mobility is highest in the center of the sensor, resulting in the peak for high fluences in Figure~\ref{fig:fig10}.  Since the local angle changes with depth, the total transverse deflection is calculated by integrating along the charge carrier trajectory.  Due to their significantly lower mobility, the impact of the Lorentz angle is negligible for holes.

\begin{figure}[htb]
\begin{center}
\includegraphics[width=0.95\textwidth]{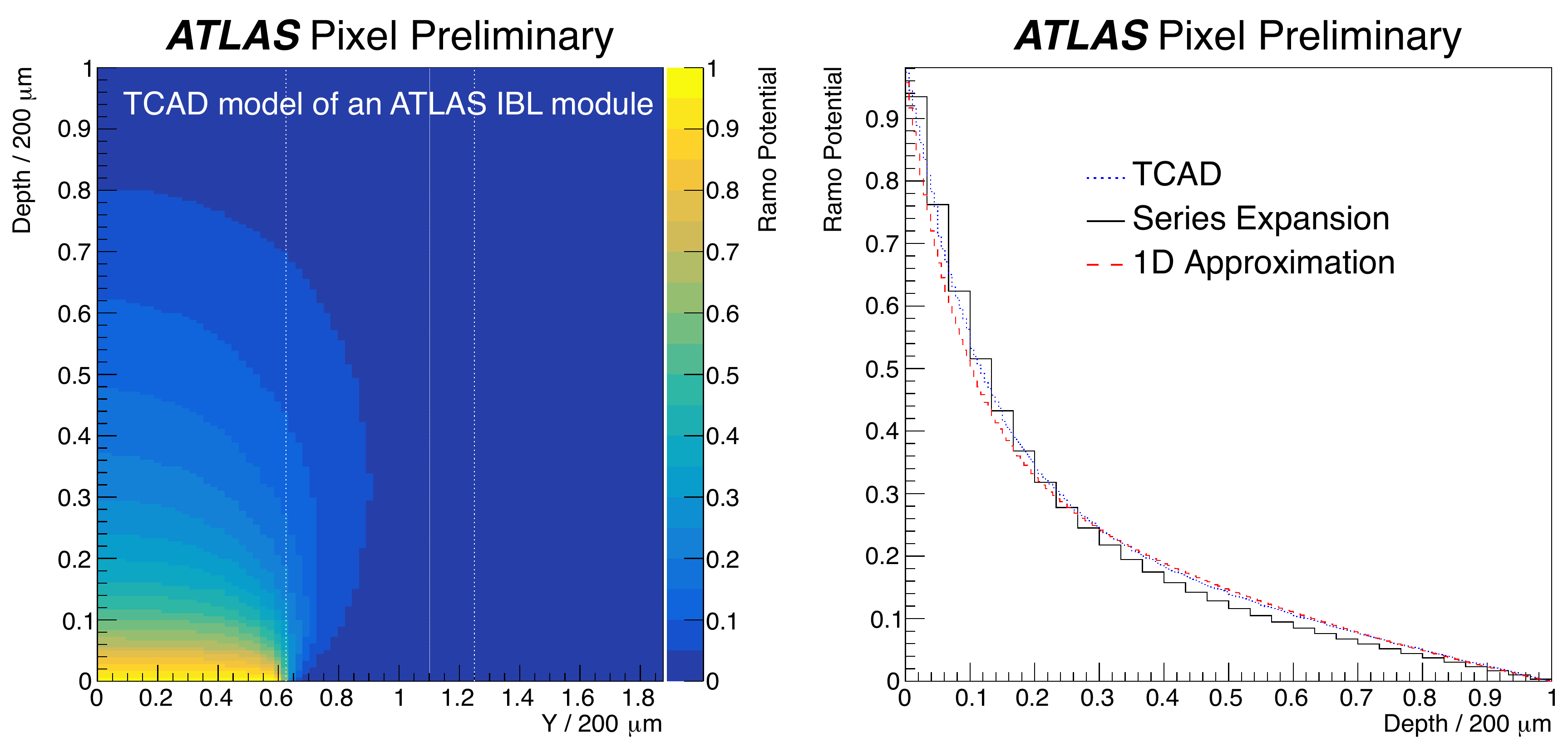}
\caption{Two- (left) and one-dimensional (right) slices of the Ramo potential inside an IBL pixel module.  Vertical white lines on the left plot are every $125$ $\mu$m.   The first white line from the left is at a pixel boundary, as $0$ is at the center of a pixel.  In the right plot, two approximations to the field are shown alongside the full numerical calculation based on TCAD.  The approximation labeled \textit{series expansion} solves the full Poisson equation using a truncated Taylor series.  The 
``1D approximation'' is a simple one-dimensional sum of exponential functions.}
\label{fig:fig12}
\end{center}
\end{figure}

\begin{figure}[htb]
\begin{center}
\includegraphics[width=0.95\textwidth]{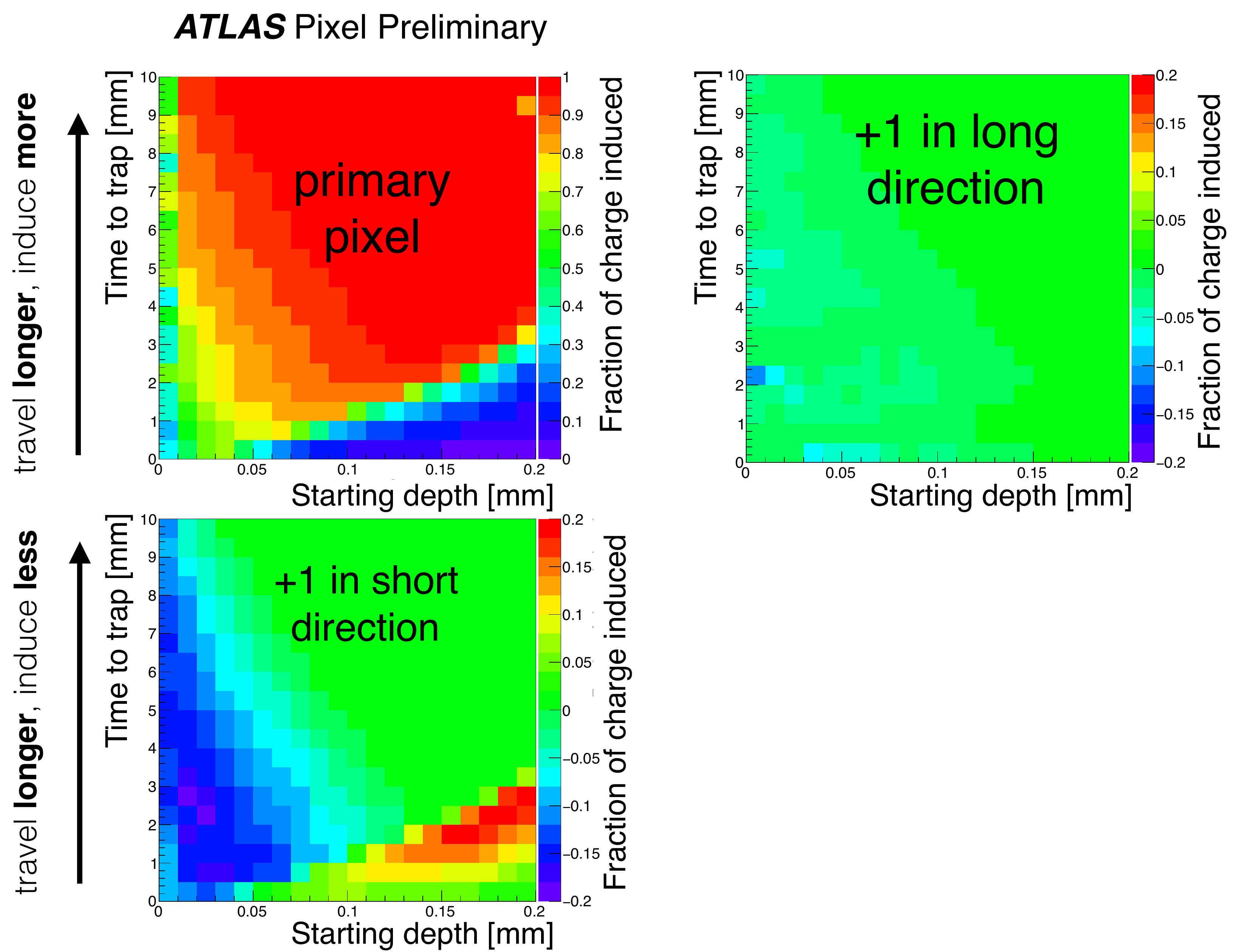}
\caption{The fraction of the deposited charge on the primary pixel that is induced on the primary pixel (top left) or the neighbor in the $+\eta$ direction (top right) or the neighbor in the $+\phi$ direction (bottom left).}
\label{fig:fig13}
\end{center}
\end{figure}

Even if charge carriers are trapped, they still contribute to the total signal via the induced charge.  As a result of the Shockley-Ramo theorem~\cite{Shockley,Ramo}, it is not necessary to keep track of the full time series of the transient current in the pixel implant.  Instead, one can compute the induced charge from the difference in the Ramo potential $\phi$ via $Q_\mathrm{induced}=e[\phi({z_\mathrm{final}})-\phi(z_0)]$.  The potential depends only on geometry and results from calculating the electric potential inside the sensor for a configuration in which the pixel implant is at $+1$ V and all other electrodes held at $0$ V.  The IBL Ramo potential is shown in Figure~\ref{fig:fig12}.  Since it decays rapidly away from the pixel implant, most of the charge is induced near $z=0$ and therefore the signal is dominated by electrons.  The potential is non-zero in neighboring pixels, so in the digitizer all of the immediate neighbors also contribute to the induced charge.  Figure~\ref{fig:fig13} demonstrates the impact of trapping on the induced charge in a particular planar sensor and its neighbors.  In the pixel the MIP traversed (primary pixel), the fraction of the deposited charge that is induced reaches 1 as the time to be trapped grows large.  Likewise, the induced charge in the neighbors approaches 0 in the same limit.  For a trapping time that is small compared to that to reach the electrode, less than the total deposited charge is induced in the primary pixel and a small amount can be measured by the neighbors.  The neighbor charge is largest in the $\phi$ direction where the pitch is smaller.  The asymmetry in the induced charge as a function of the depth results from the difference in electron and hole mobilities and the fact that electrons move toward the region of larger Ramo potential (and thus are responsible for most of the signal).


\begin{figure}[htb]
\begin{center}
\includegraphics[width=0.7\textwidth]{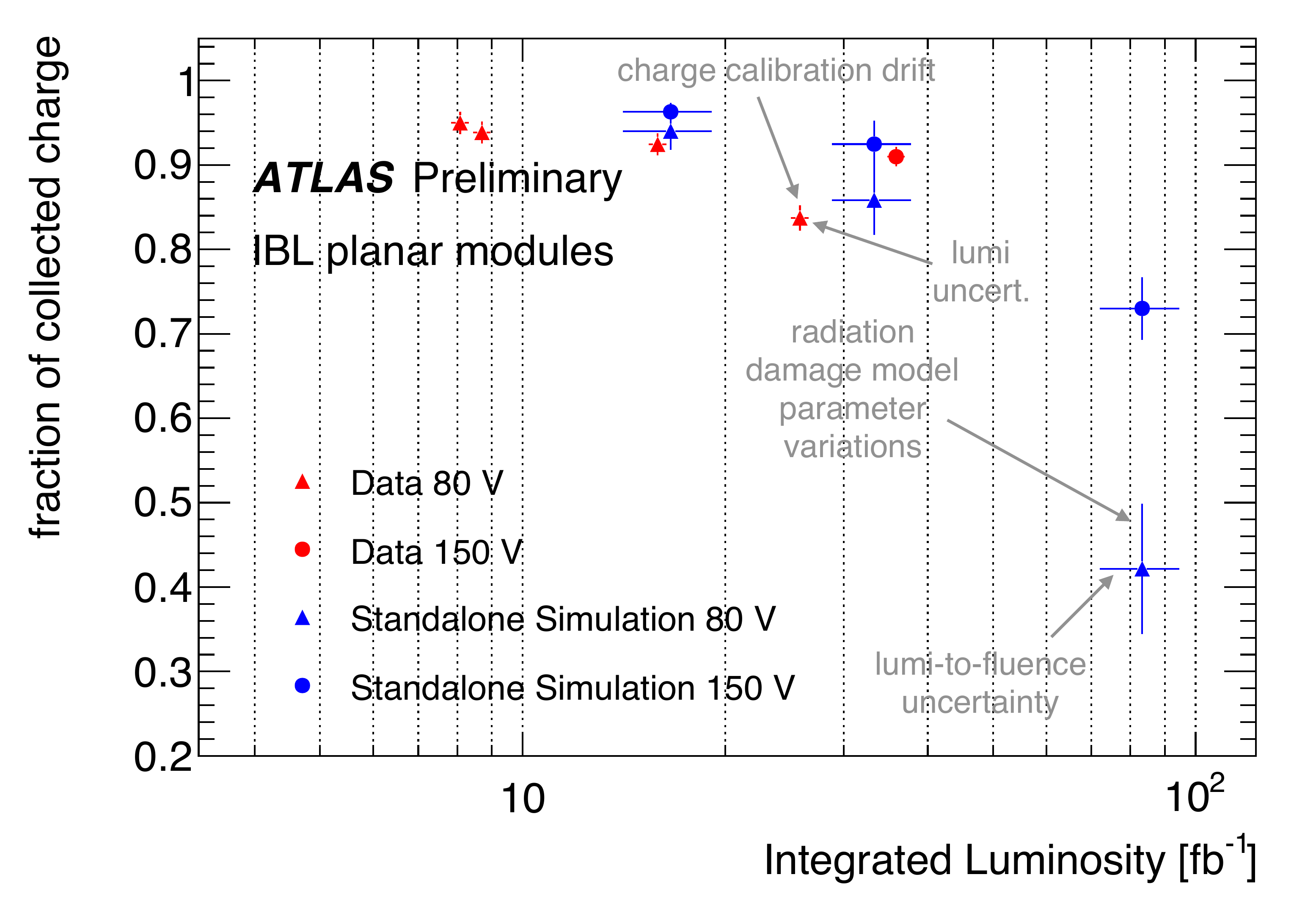}
\caption{The measured and predicted charge collection efficiency as a function of the integrated luminosity.  The charge collection efficiency is defined as the most probable value of the cluster charge distribution relative to the value at zero fuence.  Various systematic uncertainties include radiation damage model variations, integrated luminosity-to-fluence converstion factors, radiation dependence of module tunings, and the integrated luminosity uncertainty.  Adapted from Ref.~\cite{PIX-2017-004}.}
\label{fig:fig14}
\end{center}
\end{figure}

Accounting for all of the physical effects highlighted in Figure~\ref{fig:fig6} and described above\footnote{There has also been a study of annealing resulting from the short shutdown between the 2015 and 2016 runs.  The amount of annealing is predicted based on the Hamburg model (See Ref.~\cite{Moll:1999kv} and references therein) and the effective doping concentration in the TCAD simulation is adjusted accordingly.  With the present levels of annealing, this is a small effect; it will be more important for the outer layers and in the future.}, the new model can make concrete predictions for observable quantities like the charge collection efficiency shown in Figure~\ref{fig:fig14}.  Due to charge trapping and a reduction in the depletion volume, the fraction of the collected charge decreases with integrated luminosity.  After the high voltage increase (see also Figure~\ref{fig:fig3}) around $\int\mathcal{L}dt=35$ fb$^{-1}$, the efficiency increases when full depletion is restored.  Error bars on both the data and simulation reflect uncertainty in the radiation damage model, integrated luminosity-to-fluence conversion factors, and integrated luminosity, and also include the radiation dependence of module tunings.  Within these uncertainties, the simulation is consistent with the collision data recorded from the IBL.  This builds confidence in the model, allowing the extrapolation beyond the current dataset to plan for the operating voltage as well as module thresholds to tradeoff the increased LHC collision rate and the reduced signal from radiation damage.

\clearpage

\section{Conclusions and Outlook}

\begin{figure}[htb]
\begin{center}
\includegraphics[width=0.7\textwidth]{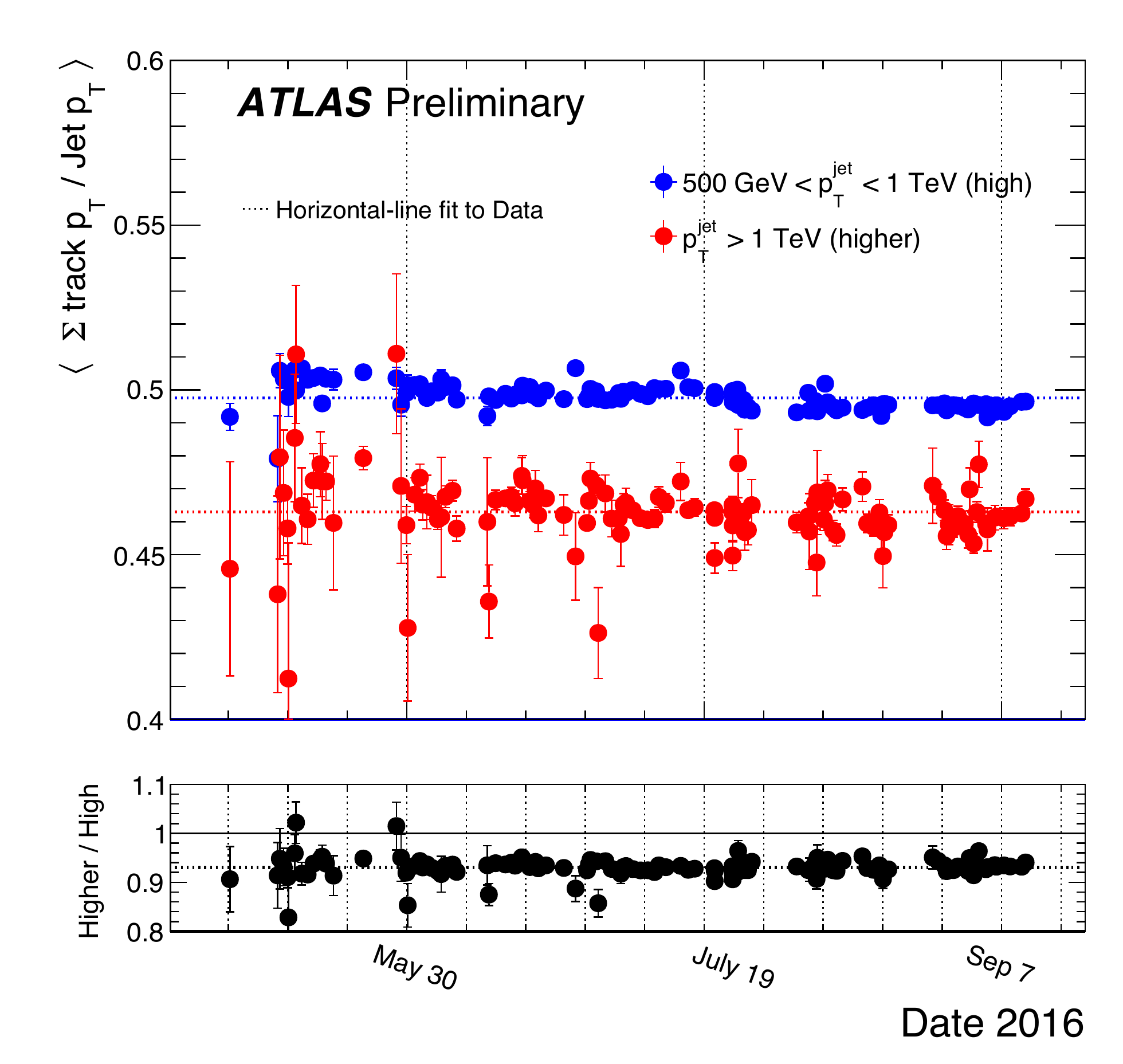}
\caption{The ratio of the sum of the track $p_\mathrm{T}$ to the calorimeter jet $p_\mathrm{T}$ inside high $p_\mathrm{T}$ jets.  At particle-level, this quantity should be $\sim 2/3$ due to the relative abundance of charged and neutral pions.  At detector level, the ratio decreases due to the track reconstruction efficiency.  It is further reduced at high $p_\mathrm{T}$ from the loss of tracks that are too close to resolve.  The ratio of the blue (high $p_\mathrm{T}$) to the red (higher $p_\mathrm{T}$) is an indication of this additional loss in efficiency.  There is no evidence for a significant degradation as a function of time in 2016.  Reproduced from Ref.~\cite{IDTR-2016-019}.}
\label{fig:fig15}
\end{center}
\end{figure}

The impact of silicon bulk radiation damage on the ATLAS Pixel sensors is already measurable with the Run 2 dataset.  The fluence levels are only $10\%$ ($1\%$) of the full (HL-)LHC projection so they will only continue to become more important.  It is therefore critical to include these effects in simulations of the ATLAS detector.  A new model has been presented, which incorporates these effects and can be used to make predictions that are consistent with collision data.  So far, tracking performance is largely unaffected due to radiation damage (see e.g. Figure~\ref{fig:fig15}), but degradation in performance is inevitable.  The new simulation can be used to study the impact of radiation damage in the future and adapt operational and offline procedures to at least partially compensate for losses.  A complete documentation of the radiation damage model is forthcoming. 


\clearpage

\Acknowledgements
This work was supported in part by the Office of High Energy Physics of the U.S. Department of Energy under contract DE-AC02-05CH11231.

\bibliographystyle{JHEP}	
\bibliography{myrefs.bib}{}
 
\end{document}